# Breakthrough of a cross-century problematic issue：The perfect statements of the third law of thermodynamics

Xiaohang Chen[+], Shanhe Su[+], Yinghui Zhou[*], Jincan Chen[*]

Department of Physics, Xiamen University, Xiamen 361005, People's Republic of China

**Abstract.** The third law of thermodynamics has been verified experimentally, but how to perfectly express such a law in theory has become a cross-century problematic issue. It is found from the recent researches that by introducing an innovative method, the Nernst equation can be obtained directly from the experimental data of chemical reactions at low temperatures without the need for artificial additional assumptions appearing in textbooks, so that the Nernst theorem should be replaced by the Nernst statement. The heat capacity statement proposed recently can be also obtained from the experimental data of the heat capacity at low temperatures. The heat capacity statement and the Nernst statement are proved to be mutually derivable and the two are equivalent. The unattainability statement of absolute zero temperature is only a coloration of the Nernst statement or the heat capacity statement. Simultaneously, the defects and deficiencies related to the contents of the third law of thermodynamics appearing in textbooks are pointed out and corrected. The results obtained show clearly that the Nernst theorem and the unattainability statement of absolute zero temperature should be withdrawn from the statements of the third law of thermodynamics. It is important to find that the Nernst statement and the heat capacity statement are two

---

[+] The two authors contributed equally to this work.
[*]Emails: yhzhou@xmu.edu.cn; jccchen@xmu.edu.cn

equivalent statements of the third law of thermodynamics, which can solve the centennial debate problems of the third law of thermodynamics and supply the perfect statements for the third law of thermodynamics.

## 1. Introduction

By the end of the nineteenth century, the theory mansion of the classical thermodynamics had been basically completed on the basis of the zeroth, first, and second laws [1-4]. However, it was found that the bottom of this beautiful building was missing a fulcrum. Although the second law of thermodynamics determines that +0 K is the lower bound of absolute temperature, it is not able to judge whether the state of the thermodynamic system at +0 K is meaningful [5]. This required a new law of thermodynamics, which stimulated the study of a large number of low-temperature experiments.

In early 20th century, many chemists had studied the chemical reactions of thermodynamic systems at low temperatures and obtained a huge amount of experimental data. Based on those data, the Nernst equation was derived in 1906，which was first called the Nernst heat theorem and later referred to as the Nernst postulate or the Nernst theorem through the work of Simon and the formulation of Planck [1]. The Nernst equation laid a foundation for the establishment of the third law of thermodynamics. However, the accurate statements of the third law of thermodynamics have not been given for a long time.

## 2. The century-old debates of the third law of thermodynamics

As a fundamental law of thermodynamics, it may have different statements, but the statements must be mutually derivable and equivalent. For example, the second law of thermodynamics has many different statements, in which two of the most common ones are the Clausius statement and the Kelvin-Planck statement [2, 3, 6]. They have been proved to be equivalent to each other in many textbooks. The third law of thermodynamics may also have different statements. The Nernst theorem and the unattainability statement (principle) of absolute zero temperature are two common statements of the third law of thermodynamics appearing in textbooks [1, 3, 6-8]. It is rather generally believed [1, 9-13] and occasionally even proved [3, 9, 10, 14] that the Nernst theorem and the unattainability statement of absolute zero temperature of the third law of thermodynamics are equivalent. However, the two were not considered to be equivalent in several textbooks [15-18] and literature [19, 20]. In fact, as early as 1920, Lewis and Gibson [21] pointed out that "the third law has been verified experimentally, but we also see some a priori reasons for the existence of such a law." Landsberg [22] argued that "the principle of the unattainability of the absolute zero does not imply Nernst's heat theorem and Nernst's heat theorem does not imply the principle of the unattainability of the absolute zero" and demonstrated the two are not equivalent. He also offered another view：The third law of thermodynamics is a collection of the Nernst theorem and the principle of the unattainability of absolute zero temperature [23]. There are even scholars [24] who believed that "the third law has been stated in several forms, no two of which are precisely equivalent, and none of which can be utilized without the introduction of some extra-thermodynamic information." The statements of the third law of thermodynamics have been a matter of debate for more than 100 years. The main debate centers on two issues:

(i) Are both the Nernst theorem and the unattainability statement of absolute zero temperature equivalent?

(ii) Is the Nernst theorem or the unattainability statement of absolute zero temperature more suitable as the statement of the third law of thermodynamics?

Although the contents and statements of the third law of thermodynamics have been a controversial issue, the derivation of the Nernst theorem in textbooks has rarely been questioned, so that the Nernst theorem as the core content of the third law of thermodynamics has been written into textbooks for more than one-century and the awkward problems caused by it have rarely been raised and discussed. As we all know, physical laws and theorems are fundamentally different, and it is rare to use a physical theorem as the statement of a physical law in the various disciplines of physics. For more than one century, even if someone asked this question, no solution could be found, so that the question silently interrogated generation after generation of researchers, and constantly tormented their minds. Therefore, the usual way to deal with this problem is to turn a blind eye, not ask, and reluctantly accept the reality. In order to bypass this problem, some scholars may directly adopt the unattainability statement of absolute zero temperature as the statement of the third law of thermodynamics, and take the Nernst theorem as a corollary of the third law of thermodynamics.

In addition to the controversy caused by the Nernst theorem and the unattainability statement of absolute zero temperature, the debate about the third law of thermodynamics also involves whether the Nernst equation and the vanishing heat capacity at absolute zero temperature are equivalent or not. In many textbooks, it has been proved that the Nernst equation can be used directly to derive an important result, that is, the heat capacity tends to zero as the temperature approaches absolute zero [3, 4], but it has not been proved that the former can be derived from the latter, and some textbooks [25] even asserted that the former cannot be derived from the latter. In 1988, it was proved for the first time that the Nernst equation may be directly derived from the vanishing heat capacity at absolute zero temperature [26]. Since then this problem has aroused some attention and discussion [27-30]. This means that whether the Nernst equation is equivalent to the vanishing heat capacity at absolute zero temperature is another controversial issue closely related to the third law of thermodynamics.

Although the statements of the third law of thermodynamics have become a world problem of great concern, few doubt that neither the Nernst theorem nor the unattainability statement of absolute zero temperature is suitably taken as the statement of the third law of thermodynamics.

The following discussion will revolve around the century-old debate problems about the third law of thermodynamics, and the results obtained will answer Blau and Halfpap's question [31]: What is the third law of thermodynamics trying to tell us?

## 3. Entropy and heat capacity of thermodynamic systems

For a general thermodynamic system, the differential equation of the internal energy $U$ is given by [1, 32-34]

$$dU = TdS + \sum_{i=1}^{n} Y_i dy_i, \qquad (1)$$

where $S$ and $T$ are, respectively, the entropy and absolute temperature of the system, $y_i$ and $Y_i$ are the generalized coordinates and corresponding generalized forces [30, 33-35], and $n$ is the number of generalized coordinates. When all the generalized coordinates $y$ remain unchanged, the heat capacity

$$C_y = T(\partial S / \partial T)_y \qquad (2)$$

may be directly derived from Eq. (1). Introducing a new thermodynamic function

$H = U - \sum_{j=1}^{l} y_j Y_j$, which may be referred to as the generalized enthalpy, we obtain the differential expression of the generalized enthalpy $H$ as [35]

$$dH = TdS + \sum_{i=l+1}^{n} Y_i dy_i - \sum_{j=1}^{l} y_j dY_j. \tag{3}$$

From Eq. (3), one obtains

$$C_x = T(\partial S / \partial T)_x, \tag{4}$$

where $x$ represents the $(n-l)$ generalized coordinates and $l$ generalized forces. When $l = 0$, $x$ represents all the generalized coordinates $y$, the third term on the right-hand side in Eq. (3) equals zero, the generalized enthalpy degenerates into the internal energy, and $C_x$ degenerates into $C_y$. When $l = n$, $x$ represents all the generalized forces $Y$, the second term on the right-hand side in Eq. (3) equals zero, and $C_x$ degenerates into $C_Y$, which is the heat capacity when all the generalized forces $Y$ are invariant. $C_x$ is the heat capacity when $x$ remains unchanged. It includes the various heat capacities of a thermodynamic system under the different constraints. For example, the heat capacity at constant volume $C_v$ and the heat capacity at constant pressure $C_P$ are, respectively, the simplest forms of $C_y$ and $C_Y$.

From Eqs. (1) and (3), one obtains

$$\left(\frac{\partial S}{\partial U}\right)_y = \frac{1}{T} = \left(\frac{\partial S}{\partial H}\right)_x, \tag{5}$$

$$\left(\frac{\partial^2 S}{\partial U^2}\right)_y = -\frac{1}{T^2 C_y}, \tag{6}$$

and

$$\left(\frac{\partial^2 S}{\partial H^2}\right)_x = -\frac{1}{T^2 C_x}. \tag{7}$$

Eq. (6) indicates that in the regime of $T > 0$, $C_y > 0$ since $(\partial^2 S / \partial U^2)_y < 0$ is the equilibrium stability condition of a general thermodynamic system [3, 26, 36, 37]. As pointed out in Refs. [38-41], the condition of stability of a thermodynamic system is the concavity of the entropy [1]. It is clearly seen from Eq. (5) that in the regime of $T>0$, both the $H \sim S$ curve and the $U \sim S$ curve have the same curved shape and the slope of each point on the two curves equals $1/T$. It implies the fact that the second partial derivatives of the entropy $S$ with respect to $U$ and $H$ have the same sign, i.e., $(\partial^2 S/\partial U^2)_y < 0$ and $(\partial^2 S/\partial H^2)_x < 0$. Thus, it requires that in the regime of $T>0$,

$$C_x > 0. \tag{8}$$

According to Eq. (4), we obtain the entropy of a general thermodynamic system as [3, 42]

$$S = S(T,x) = S(T_0,x) + \int_{T_0}^{T} (C_x/T)dT, \tag{9}$$

where $S(T_0,x)$ is the entropy of the system at $T_0$. Eq. (9) establishes the connection between the entropy and the heat capacity of a thermodynamic system.

## 4. The Nernst theorem

In the late 19th century and early 20th century, it was found from an enormous amount of the experimental data at low-temperature chemical reactions that the changes $\Delta H$ and $\Delta G$ of the enthalpy $H$ and Gibbs free energy $G$ of thermodynamic systems in the isothermal and isobaric process become closer and closer as the temperature decreases. When the temperature is extrapolated to absolute zero [1], there is $(\Delta G)_0 = (\Delta H)_0$, as shown in Fig. 1(a) [1, 6, 43-45].

In some textbooks [42, 46], it is common to define the chemical affinity $A = -\Delta G$ of a system in the isothermal isobaric process and the heat $Q = -\Delta H$ released by the system. $Q$

is conventionally called the heat of reaction because $Q > 0$ [46] is the more common case for most chemical reactions [42]. By applying the experimental data of $\Delta G$ and $\Delta H$, a schematic diagram of $A$ and $Q$ in the isothermal and isobaric process varying with $T$ can be plotted, as shown in Fig. 1 (b) [46, 47].

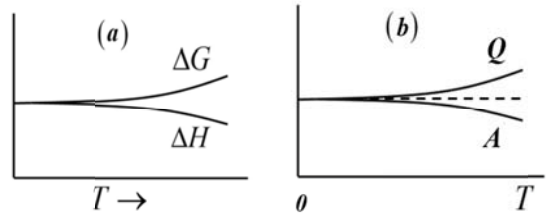

Fig.1. The conventional schematic diagrams of (a) $\Delta H$ and $\Delta G$ and (b) $Q$ and $A$ of the thermodynamic systems in the isothermal and isobaric process varying with $T$.

According to the definition of the Gibbs free energy

$$G = H - TS, \tag{10}$$

the fundamental relation between $\Delta H$ and $\Delta G$ of a thermodynamic system in the isothermal process is given by

$$\Delta G = \Delta H - T\Delta S, \tag{11}$$

where $\Delta S$ is the entropy change of a thermodynamic system in an isothermal process. Using two artificial additional assumptions typically introduced in textbooks [1, 6], i.e., $\Delta S$ is bounded when $T \to 0$ and $\lim_{T \to 0}(\partial \Delta H / \partial T) = \lim_{T \to 0}(\partial \Delta G / \partial T)$, one can derive the Nernst equation, i.e.,

$$\lim_{T \to 0}(\Delta S)_T = 0. \tag{12}$$

Because the additional assumptions are introduced in the derivative process, the Nernst equation is also called the Nernst postulate or Nernst theorem [1, 6, 26], which is taken as the core contents of the third law of thermodynamics. It shows clearly that the Nernst equation is very important in the theory of thermodynamics [1, 3, 4, 14, 48]. However, some questions related to the Nernst equation are rarely mentioned and discussed. The first question is that there are different versions of the schematic diagram related to the Nernst equation in textbooks [1, 6, 42, 43, 45, 46] and literature [44, 47] and are they correct? The second question is whether there are other meaningful schematic diagrams related to the Nernst equation, which have not yet been given [49]. The third question is whether the Nernst equation can be re-obtained from the experimental data of low-temperature chemical reactions without any artificial additional assumptions. The fourth question is whether the awkward problem caused by the Nernst theorem with artificial additional assumptions used as the core contents of the third law of thermodynamics can be avoided. These questions are worthy of special discussion.

**5. The schematic diagrams of the Nernst equation.**

Considering the fact that $\Delta H$ and $\Delta G$ are the experimental data obtained from the thermodynamic systems in the isothermal and isobaric process, one can determine $\Delta G < 0$ when $T > 0$, because the irreversible process that occurs in the isothermal and isobaric system always proceeds in the direction of the reduction of the Gibbs free energy [6, 8]. There are two cases for $\Delta H$, i.e., $\Delta H > 0$ and $\Delta H < 0$, which will be, respectively, discussed below.

(i) The case of $\Delta H > 0$. In the region of $T > 0$, $\Delta H > 0 > \Delta G$. When the experimental data of low temperature chemical reactions are extrapolated to absolute zero, there is $(\Delta G)_0 = (\Delta H)_0$. Based on the two conditions mentioned just, it is necessary to have $(\Delta H)_0 = 0$. Thus, the curves of $\Delta H$ and $\Delta G$ varying with $T$ should be schematically shown in Fig. 2 (a).

(ii) The case of $\Delta H < 0$. In the temperature range of $T > 0$, $\Delta H = \Delta Q < 0$ for a simple thermodynamic system in the isothermal and isobaric process, where $\Delta Q$ indicates the heat absorbed by the system. $\Delta Q < 0$ represents that the system releases heat to the environment in the isothermal isobaric process. For a reversible isothermal exothermic process, $\Delta S < 0$ can be directly obtained from $\Delta Q = T\Delta S < 0$. For an irreversible isothermal exothermic process, $\Delta Q < T\Delta S$. It cannot be judged from both $\Delta Q < 0$ and $\Delta Q < T\Delta S$ whether $\Delta S$ is smaller than 0 or not. The entropy change $\Delta S$ of the system in an irreversible isothermal exothermic process is composed of the entropy decrease caused by the heat released in the isothermal process and the entropy increase caused by the irreversibility inside the system. $\Delta S < 0$ holds only when the entropy change caused by isothermal heat release is dominant. It can be seen from Eq. (11) that in such a case, $0 > \Delta G > \Delta H$. When $T \to 0$, $(\Delta G)_0 = (\Delta H)_0$. The curves depicting the variations of $\Delta G$ and $\Delta H$ with respect to $T$ should be schematically illustrated in Fig. 2(b). When the entropy change caused by the irreversibility inside the system is dominant, $\Delta S > 0$. It be seen from Eq. (11) that $0 > \Delta H > \Delta G$. When $T \to 0$, $(\Delta G)_0 = (\Delta H)_0$. The curves of $\Delta G$ and $\Delta H$ varying with $T$ should be schematically illustrated in Fig. 2(c). When the entropy change caused by the isothermal heat release is equal to that caused by the irreversibility inside the system, $\Delta S = 0$, as shown in Fig. 2(d). This is exactly the case what textbook [1] describes: the enthalpy change $\Delta H$ is very nearly equal to the Gibbs free energy change $\Delta G$ over a considerable temperature range.

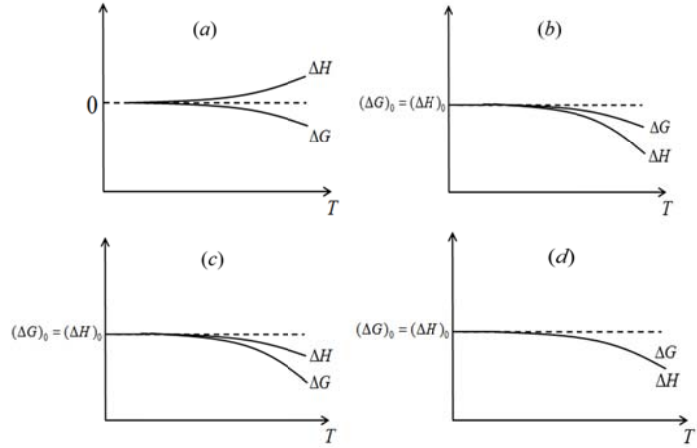

Fig. 2. The schematic diagrams of $\Delta H$ and $\Delta G$ of the thermodynamic systems in the isothermal and isobaric process varying with $T$, where (a) $\Delta H > 0 > \Delta G$ and $(\Delta G)_0 = (\Delta H)_0 = 0$, (b) $0 > \Delta G > \Delta H$ and $(\Delta G)_0 = (\Delta H)_0$, (c) $0 > \Delta H > \Delta G$ and $(\Delta G)_0 = (\Delta H)_0$, and (d) $0 > \Delta H = \Delta G$ and $(\Delta G)_0 = (\Delta H)_0$.

If the system has undergone sufficient evolution prior to each test, the chemical reaction is basically over, the system approaches the equilibrium state, the isothermal and isobaric processes tend to be reversible in the region of $T > 0$, and $\Delta Q = T \Delta S$. In such a case, $\Delta G = 0$. When $T \to 0$, $(\Delta G)_0 = (\Delta H)_0 = 0$. For the case of reversible processes, Figs. 2(c) and (d) no longer exist, and Figs. 2(a) and (b) are reduced to Figs. 3 (a) and (b), respectively.

It can be clearly seen from Eq. (10) that the isothermal isobaric process mentioned above is only a sufficient condition that Eq. (11) is true, while the isothermal process is a necessary condition that Eq. (11) is true. Now, we continue to discuss that the cases of $\Delta H$ and $\Delta G$ of the thermodynamic system in an isothermal process varying with $T$.

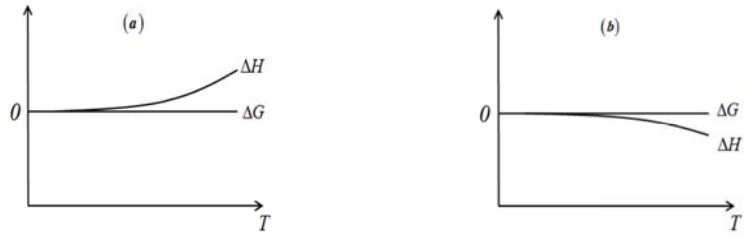

Fig. 3. The schematic diagrams of $\Delta H$ and $\Delta G$ of the thermodynamic systems in the revisable isothermal and isobaric process varying with $T$, where (a) $\Delta H > 0 = \Delta G$ and $(\Delta G)_0 = (\Delta H)_0 = 0$, and (b) $0 = \Delta G > \Delta H$ and $(\Delta G)_0 = (\Delta H)_0 = 0$.

For a simple thermodynamic system only including volume variable work, the enthalpy change in the isothermal process can be expressed as

$$\Delta H = \Delta Q + v\Delta P, \qquad (13)$$

where the change of the Gibbs free energy is still represented by Eq. (11), $\Delta Q$ represents the heat absorbed by the system from the environment during the isothermal process, and $v$ and $P$ are the volume and pressure of the system. When the isothermal process is irreversible, $\Delta Q < T\Delta S$. According to Eqs. (11) and (13), $\Delta G < v\Delta P$. In the region of $T > 0$, when $\Delta H > v\Delta P$, there is $\Delta H > v\Delta P > \Delta G$; When $T \to 0$, $(\Delta H)_0 = (\Delta G)_0 = (v\Delta P)_0$; as shown in Fig. 4 (a). When $\Delta H < v\Delta P$, $\Delta Q < 0$ may be determined by Eq. (13). This means that the system releases heat to the environment. In the region of $T > 0$, $\Delta Q < 0$ and $\Delta Q < T\Delta S$ cannot determine whether $\Delta S$ is less than zero. When $\Delta S < 0$, $(v\Delta P) > \Delta G > \Delta H$; When $\Delta S > 0$, $(v\Delta P) > \Delta H > \Delta G$; When $\Delta S = 0$, $(v\Delta P) > \Delta G = \Delta H$; as shown in Figs. 4 (b) - (d), respectively. When $T \to 0$, $(\Delta H)_0 = (\Delta G)_0 = (\Delta Q)_0 + (v\Delta P)_0$.

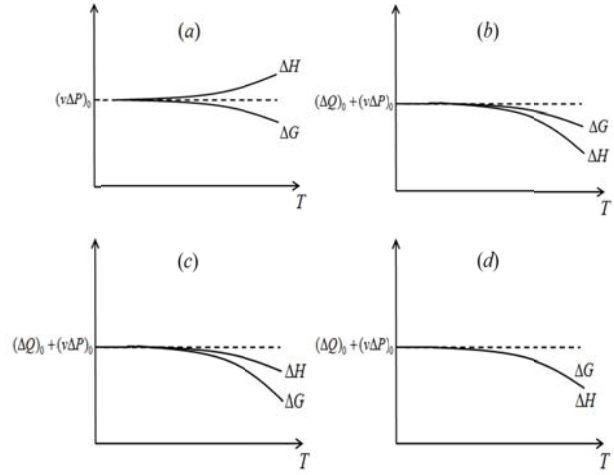

Fig. 4. The schematic diagrams of $\Delta H$ and $\Delta G$ of the thermodynamic systems in the isothermal process varying with $T$, (a) $\Delta H > (v\Delta P) > \Delta G$, (b) $(v\Delta P) > \Delta G > \Delta H$, (c) $(v\Delta P) > \Delta H > \Delta G$, and (d) $(v\Delta P) > \Delta H = \Delta G$. When $T \to 0$, (a) $(\Delta H)_0 = (\Delta G)_0 = (v\Delta P)_0$, and (b)-(d) $(\Delta H)_0 = (\Delta G)_0 = (\Delta Q)_0 + (v\Delta P)_0$.

When the isothermal process is revisable, $\Delta Q = T\Delta S$. It can be seen from Eqs. (11) and (13) that Figs. 4 (c) and (d) do not exist, while Figs. 4 (a) and (b) are simplified to Figs. 5 (a) and (b), respectively. In Fig. 5, for simplicity, $v\Delta P$ is assumed to be independent of temperature and is a constant.

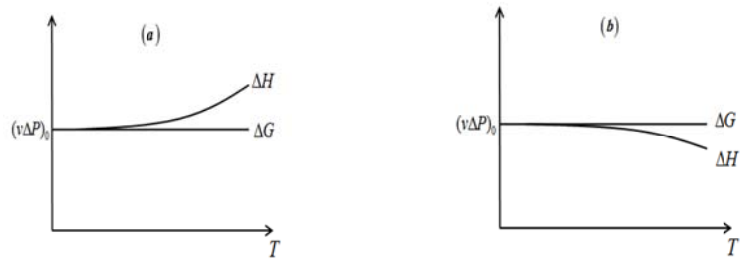

Fig. 5. The schematic diagrams of $\Delta H$ and $\Delta G$ of the thermodynamic systems in the revisable isothermal process varying with $T$. In the region of $T > 0$, (a) $\Delta H > (v\Delta P) = \Delta G$ and (b) $(v\Delta P) = \Delta G > \Delta H$. When $T \to 0$, $(\Delta H)_0 = (\Delta G)_0 = (v\Delta P)_0$.

According to Figs. 2-5 and the definitions of $A$ and $Q$, one can easily generate the curves of $Q$ and $A$ varying with $T$. For example, the schematic diagrams of $Q$ and $A$ of the thermodynamic systems in the isothermal and isobaric process varying with $T$ are shown in Fig. 6.

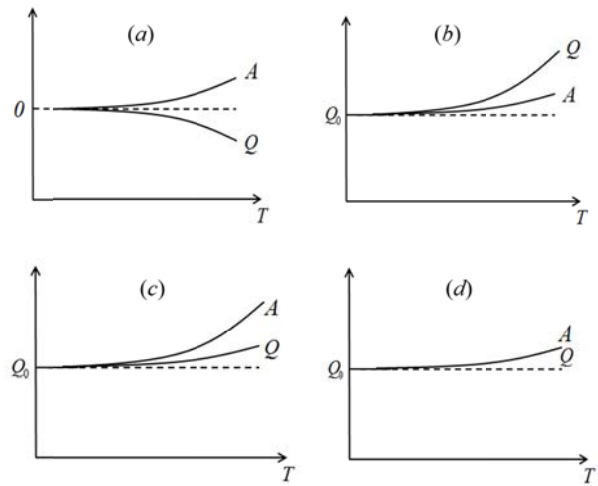

Fig.6. The schematic diagrams of $Q$ and $A$ of the thermodynamic systems in the isothermal and isobaric process varying with $T$, where (a) $A > 0 > Q$ and $A_0 = Q_0 = 0$, (b) $Q > A > 0$ and $A_0 = Q_0$, (c) $A > Q > 0$ and $A_0 = Q_0$, and (d) $A = Q > 0$ and $A_0 = Q_0$.

According to Figs. 2-6, Eq. (11), and two artificial additional assumptions mentioned above, one can conveniently derive Eq. (12).

For an isothermal process of a thermodynamic system, $\Delta Q \leq T \Delta S$. When the system is surrounded by an environment much larger than it, both of which are at absolute zero temperature, $\Delta Q \leq 0$ as long as Eq. (12) holds. This means that during the isothermal process of $T = 0$, endothermic heat does not occur. Similarly, the heat absorption of the environment at $T = 0$ will also not occur. If the isothermal process of the thermodynamic system is irreversible, the system must release heat, but the environment will not absorb heat, so that the irreversible isothermal process of the thermodynamic system cannot be carried out, as shown in Fig.7. This shows that the isothermal process of a thermodynamic system at $T=0$ must be reversible as long as Eq. (12) holds. As described in textbooks, the isothermal process carried out by the thermodynamic system at $T=0$ coincides with the reversible adiabatic process [1, 8], as shown in Fig.8, and there is no heat exchange between the system and the environment. Thus, $(\Delta H)_0$ in Figs. 2(b)-(d) is equal to zero and $Q_0$ in Figs. 6(b)-(d) is also equal to zero. However, for the thermodynamic systems in the isothermal process rather than the isothermal and isobaric process, $(v\Delta P)_0$ in Figs.4 and 5 is not required to equal to zero so that $(\Delta H)_0$ may not be equal to zero. This energy $(v\Delta P)_0$ translates as the internal energy of the system. When $(v\Delta P)_0 < 0$, the system does work on the environment.

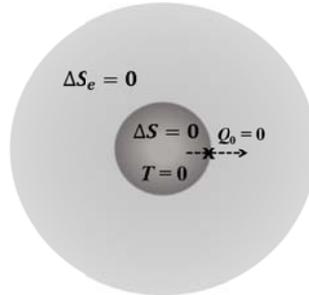

Fig.7. The schematic diagram of a thermodynamics system at $T=0$.

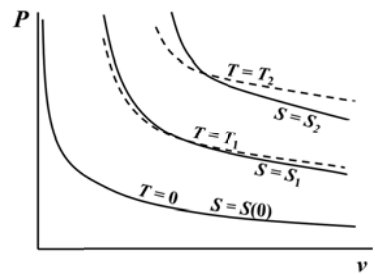

Fig.8. The schematic diagram of a thermodynamic system changing from $T>0$ to $T=0$.

It can be found to compare Figs.2 and 6 with Fig.1 that Figs. 1(a) and (b) are incorrect, because the schematic diagrams in Fig.1 do not satisfy the basic requirement of $\Delta H > 0 > \Delta G$ or ($0 > \Delta G, 0 > \Delta H$) in the region of $T > 0$. The errors in Figure 1 may have originated in Ref. [47] or in other previous references. Unfortunately, these wrong diagrams have been ignored for a long time and adopted by most textbooks [1, 6, 43-46] for more than one century. It can be also found that Figs. 2 (b)-(d) and 6 (b)-(d) never appear in textbooks and literature and are the new schematic diagrams related to the Nernst equation, which should be paid attention to in teaching and textbook compilation of universities. Figs.3-5 may be some meaning schematic diagrams related to the Nernst equation, but these diagrams have never been seen in history. So far we have answered the first and second problems raised in the preceding section.

## 6. The Nernst statement

It can be seen from Figs. 2-5 that when the curves of $\Delta H$ and $\Delta G$ of a thermodynamic system varying with $T$ are plotted, it is not only necessary to indicate whether the system is in an isothermal isobaric process or an isothermal process, but also whether the process is reversible or irreversible. It shows clearly that even for a same thermodynamic system, the schematic diagrams of the Nernst equation are different for different experimental test conditions. It can be also seen from Figs. 2-5 that these schematic diagrams are suitable for the thermodynamic systems only including volume variable work. It means that for different thermodynamic systems, the schematic diagrams of the Nernst equation are different. In particular, for general thermodynamic systems described by Eq. (1), it is highly meaningful to know how to generate the schematic diagrams of the Nernst equation.

It was found from an amount of experimental data at low temperatures that $\Delta H$ of a thermodynamic system at different experimental test conditions are very different and so are $\Delta G$. However, they have a common feature: no matter whether the thermodynamic system is a simple system or a general system and whether the system is in an isothermal isobaric process or an isothermal process, the experimental data of $\Delta H$ and $\Delta G$ become closer and closer as the temperature decreases. When $T \to 0$, $(\Delta G)_0 = (\Delta H)_0$. With the help of this feature, one can innovatively introduce a new function $f \equiv |\Delta H - \Delta G|$ and easily generate the curves of $f$ varying with $T$, as indicated in Fig.9. The importance of Fig.9 lies in the fact that the schematic diagram of the Nernst equation in any thermodynamic system in the isothermal process and isothermal isobaric process can be generated in a unified way whether $\Delta H > 0$ or $\Delta H < 0$. It shows that Fig.9 is suitable for not only a simple thermodynamic system but also a general thermodynamic system [50].

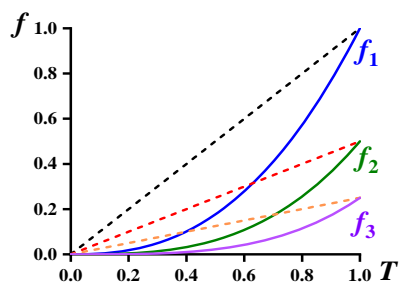

Fig. 9. The schematic diagram of the new function $f$ of the thermodynamic systems in the isothermal process varying with $T$.

It is seen from Fig.9 that $f_i$ $(i = 1, 2, ...)$ is a function of $T$. For different thermodynamic systems, $f_i$ may have different forms. It is found from a large number of experimental data that for general thermodynamic systems at ultra-low temperatures, $f_i$ has one common feature: The decrease of $f_i$ with temperature is faster than that of $T$ itself [50, 51]. The general expression of $f_i$ may be given by

$$f_i = a_{i1}T^\alpha + a_{i2}T^\beta + a_{i3}T^\gamma + ..., \tag{14}$$

where $a_{ij} \geq 0$ ($j = 1, 2, 3, ...$) are some coefficients to be independent of temperature, and $1 < \alpha < \beta < \gamma$. The simplest form of $f_i$ is $f_i = a_{i1}T^\alpha$.

It is seen from the curves of $f_i$ varying with $T$ that the lower the temperature is, the smaller the slope of curves, resulting in a result that the entropy change $(\Delta S)_T$ of thermodynamic systems during the isothermal process becomes smaller. When $T \to 0$, the slope of curves approaches zero and $(\Delta S)_T = 0$. Moreover, substituting $f_i$ into Eq. (11) and extrapolating it to absolute zero temperature, one can determine

$$\lim_{T \to 0} \frac{|\Delta H - \Delta G|}{T} = \lim_{T \to 0} \frac{f_i}{T} = \lim_{T \to 0} |(\Delta S)_T| = 0 = \lim_{T \to 0} (\Delta S)_T. \tag{15}$$

The last equation in Eq. (15) is exactly the Nernst equation, i.e., Eq. (12). It should be emphasized here that without any artificial additional assumptions, the Nernst equation can be directly obtained from the experimental data of thermodynamic systems at low-temperature chemical reactions. This is obviously different from the methods in textbooks, where some artificial additional assumptions were introduced. Thus, the physical contents included in the Nernst equation should not be referred to as the Nernst postulate or the Nernst theorem [1-4, 6-15] in textbooks and literature. The relevant contents of the Nernst equation in textbooks should be rewritten. The conventional terminologies, such as the Nernst postulate or the Nernst theorem, prevalent in textbooks and literature, should be expunged and aptly renamed as the Nernst statement. Hitherto, we have solved the third and fourth problems raised in Sect.4.

There are usually two different views on the Nernst equation in textbooks. The first view [6, 24, 42] is that the entropy change associated with any isothermal process of a condensed system approaches zero as the temperature approaches absolute zero, and the second view [1, 4, 46] is that the entropy change associated with any isothermal reversible process of a condensed system approaches zero as the temperature approaches absolute zero. It can be clearly seen from Figs. 2-6 and 9 that the experimental measurements are usually completed in the irreversible process in the region of $T > 0$. When the temperature is extrapolated to absolute zero, both the isothermal isobaric process and the isothermal process in Figs. 2-6 and 9 tend to be reversible. Thus, the second view is more accurate than the first and is acceptable. However, it does not contain the characteristics of irreversible isothermal processes and entropy changes of the thermodynamic system in the region of $T > 0$. Therefore, the physical connotation contained in the Nernst equation can be interpreted as the entropy change of any thermodynamic system in the isothermal process decreases with the decrease of temperature. When the absolute temperature tends to zero, the isothermal process tends to be reversible and the entropy change of the system approaches zero. Such a statement may be called the Nernst statement and is more comprehensive and accurate than that given in textbooks, not only including the performance changes of thermodynamic systems at low temperatures but also revealing the characteristics of thermodynamic systems at $T = 0$.

**7. The heat capacity statement**

As mentioned above, besides the Nernst equation can be obtained directly from the experimental data of chemical reactions at low temperatures, the vanishing heat capacity at absolute zero temperature can be also obtained directly from the experimental data of heat capacity at low temperatures, which may be referred to as the heat capacity statement and stated as follows [50, 52]: The various heat capacities of a thermodynamic system under different constraints approach zero as $T \to 0$, i.e.,

$$\lim_{T \to 0} C_x = 0. \tag{16}$$

It is worth pointing out that $C_x$ represents the various heat capacities of the real thermodynamic systems at ultra-low temperatures, but does not include the fictional heat capacities [23, 53] impossibly appearing in real thermodynamic systems at ultra-low temperatures [54]. $C_x$ has a common characteristic: It decreases with the decrease of $T$. With the development of cryogenic technology, the heat capacities of many thermodynamic systems at ultra-low temperatures such as quantum Fermi gas, electron gas in metal, quantum Bose gas, superfluid, etc. have been experimentally measured. When those experimental data are simulated and extrapolated to absolute zero temperature, one can obtain Eq. (16). For example, according to Refs. [55-57], we can obtain the experimental data of the heat capacity of three materials at different temperatures and simulate the curves of $C$ varying with $T$, as shown in Fig. 10, where the curve equations can be summarized as

$$C = aT + bT^3, \tag{17}$$

where $a$ and $b$ are two constants to be independent of temperature. Eq. (17) is only a special form of the heat capacity expressions of thermodynamic systems at ultra-low temperatures. It can be found from a part of the experimental data measured in the last decades that the specific heat of many materials varying with temperature is described by the following simulation function [58-64]

$$C = aT^\lambda + bT^\mu, \qquad (18)$$

where $0 < \lambda < \mu$.

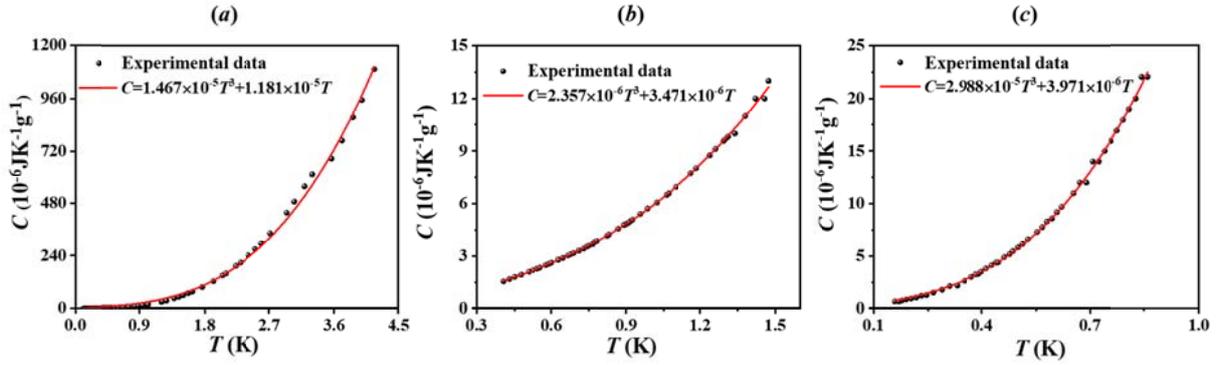

Fig. 10. The experimental data and simulation curves of the specific heat of (a) Indium, (b) Gold, and (c) Torlon varying with temperature.

Besides Eqs. (17) and (18), the varying relations of the heat capacity of thermodynamic systems at ultra-low temperatures with temperature can have a variety of other different forms [54, 65-67]. It can be found from Eqs. (17) and (18) and experimental data of the heat capacity of thermodynamic systems at ultra-low temperatures that the various heat capacities $C_x$ of thermodynamic systems at ultra-low temperatures have a common feature: $C_x$ decreases with the decrease of temperature and tends to zero when $T \to 0$, i.e., Eq. (16). It is important to note the fact that Eq. (16) is obtained by extrapolating the simulation functions from the experimental data of the specific heat of many materials to absolute zero temperature and does not need any additional assumptions. It's worth pointing out that the many expressions of heat capacity do not satisfy Eq. (16), such as the heat capacities of ideal gases [5] and magnetic materials obeying the Curie law [3, 42]. There are also some thermodynamic systems, whose heat capacities are not of monotonic functions of temperature. For example, glass is a two-level system and has a heat capacity peak within a certain temperature range as the temperature varies [68]. Such heat capacities may be suitable for thermodynamic systems in other temperature regions but impossibly appear in thermodynamic systems at ultra-low temperatures.

## 8. The equivalent proofs of two statements

By looking at the above two statements only from the experiment, they appear to be unrelated to each other. It is significant to note that there exist certain connections between the two statements because they correspond to the entropy and heat capacity of a system, respectively. Using the theory of thermodynamics, we will explain clearly the interrelationships of the two statements mentioned above.

(i) The first proof method

The entropy of a thermodynamic system is given by $S = S(T, x)$ and its differential is expressed as [69]

$$dS = (\partial S / \partial T)_x dT + \sum_{k=1}^{n} (\partial S / \partial x_k)_{T, x_l \neq x_k} dx_k. \tag{19}$$

When a thermodynamic system starts from the state $\{T, x\}$ and attains the state $\{T + \Delta T, x + \Delta x\}$, the entropy change of the system in such a process is equal to $\Delta S = S(T + \Delta T, x + \Delta x) - S(T, x)$, which is the sum of the entropy change $(\Delta S)_x$ caused by a change of $T$ when $x$ remains constant and the entropy change $(\Delta S)_T$ caused by a change of $x$ when $T$ remains constant, i.e.,

$$\Delta S=(\Delta S)_x+(\Delta S)_T. \tag{20}$$

It has been emphasized in textbooks [3, 46, 70] that no matter how low the temperature $T$ is, one can use a reversible adiabatic process to lower the temperature of a system still further. When the process is reversible adiabatic, $\Delta S = 0$, and the temperature change of the system is $(\Delta T)_S$. In such a process, $(\Delta T)_S$ may be controlled to be small enough, so that $(\Delta S)_x = (\partial S/\partial T)_x (\Delta T)_S = (C_x/T)(\Delta T)_S$. Thus, we have

$$(\Delta S)_T = -(\Delta S)_x = \frac{-(\Delta T)_S}{T} C_x. \tag{21}$$

When $T \to 0$, Eq. (21) can be expressed as

$$\lim_{T \to 0} \frac{-(\Delta T)_S}{T} = \lim_{T \to 0} \frac{(\Delta S)_T}{C_x}. \tag{22}$$

For any adiabatic cooling process starting from an arbitrary low temperature $T$, $-(\Delta T)_S$ is impossibly larger than $T$ and $-(\Delta T)_S/T$ always is a finite value larger than zero [30, 46] because +0 K is the lower limit of the absolute temperature [2, 3, 5, 46, 71]. Such a conclusion does not involve the unattainability statement of absolute zero temperature.

It is easily observed that using Eqs. (12) and (22), we can directly determine that $\lim_{T \to 0}(0/C_x)$ is equal to a finite value and mathematically derive Eq. (16), i.e., the heat capacity statement. It is also observed that that using Eqs. (16) and (22), we can directly determine that $\lim_{T \to 0}[(\Delta S)_T/0]$ is equal to a finite value and mathematically derive Eq. (12). It is very meaningful that in such a concise proof process [69], no other additional conditions are needed. By the way, the conclusions in Ref. [30] can be directly derived from those obtained here because $C_x$ includes $C_y$.

(ii) The second proof method

Using Eqs. (3) and (10), we can obtain

$$dG = -SdT + \sum_{i=l+1}^{n} Y_i dy_i - \sum_{j=1}^{l} y_j dY_j \qquad (23)$$

and

$$S = -\left(\frac{\partial G}{\partial T}\right)_x = -\frac{G-H}{T}. \qquad (24)$$

When the temperature is extrapolated to absolute zero and Eq. (12) is used, Eq. (24) may be expressed as

$$\lim_{T \to 0} S = S_0 = -\lim_{T \to 0}\left(\frac{\partial G}{\partial T}\right)_x = -\lim_{T \to 0}\frac{G-H}{T} = -[\lim_{T \to 0}\left(\frac{\partial G}{\partial T}\right)_x - \lim_{T \to 0}\left(\frac{\partial H}{\partial T}\right)_x]. \qquad (25)$$

From Eqs. (3), (4), and (25), we obtain

$$\lim_{T \to 0}\left(\frac{\partial H}{\partial T}\right)_x = \lim_{T \to 0} T\left(\frac{\partial S}{\partial T}\right)_x = \lim_{T \to 0} C_x = 0, \qquad (26)$$

which is accurately Eq. (16). This shows clearly that Eq. (16) is a direct consequence of the Nernst equation and a universal result of thermodynamics.

When $T_0 = 0$, substituting Eq. (16) into Eq. (9) yields [3, 42]

$$S(T,x) = S(0,x) + \int_0^T (C_x/T)dT. \qquad (27)$$

where $S(0,x)$ is the entropy of the system at $T_0 = 0$. Mathematically, there are no additional requirements for Eq. (27). Physically, some additional requirements are necessary for Eq. (27), because the entropy of a thermodynamic system always is finite, which requires that Eq. (16) must be satisfied and the integral term $\int_0^T (C_x/T)dT$ in Eq. (27) is a finite value. In other words, the heat capacitors $C_x$ of a real thermodynamic system at ultra-low temperatures can ensure the integral term $\int_0^T (C_x/T)dT$ in Eq. (27) to be a finite value.

It is worth pointing out that Eq. (27) only includes Eqs. (9) and (16), and consequently, it can be directly used to derive the Nernst equation. According to Eq. (16), we can prove that for a real thermodynamic system at ultra-low temperatures,

$$0 < \int_0^{T_j} [C_x(x,T)/T]dT \equiv F(x,T_j). \tag{28}$$

For a thermodynamic system evolving along a quasistatic reversible adiabatic process, $x$ changes from $x'$ to $x''$, and $T$ changes from $T_1$ to $T_2$, whereas the entropy $S$ remains unchanged. Using Eqs. (27) and (28), we can obtain [35,72]

$$S(0,x'') - S(0,x') = F(x',T_1) - F(x'',T_2). \tag{29}$$

It is seen from Eq. (29) that if $S(0,x'') > S(0,x')$ is assumed, one can find a positive $T_1$ to ensure that $F(x',T_1)$ is small enough and $S(0,x'') - S(0,x') > F(x',T_1)$. In such a case, it requires that $F(x'',T_2)$ must be negative, which is in contradiction with Eq. (28); similarly, if $S(0,x'') < S(0,x')$ is assumed, one can find a positive $T_2$ to ensure that $F(x'',T_2)$ is small enough and $S(0,x') - S(0,x'') > F(x'',T_2)$, and consequently, it requires that $F(x',T_1)$ must be negative, which is also in contradiction with Eq. (28). Thus, we must have a conclusion that $S(0,x'') = S(0,x')$ as $T \to 0$, which is accurately the Nernst equation.

(iii) The third proof method

According to the Nernst equation, we have

$$\lim_{T \to 0} S = S_0 = \lim_{T \to 0} \frac{TS}{T} = \lim_{T \to 0} [\frac{\partial (TS)}{\partial T}]_x = \lim_{T \to 0} S + \lim_{T \to 0} (T \frac{\partial S}{\partial T})_x \tag{30}$$

From Eq. (30), one can obtain

$$\lim_{T \to 0} (T \frac{\partial S}{\partial T})_x = \lim_{T \to 0} C_x = 0. \tag{31}$$

It shows clearly that Eq. (16) can be directly from Eq. (12).

If starting from Eq. (31), one can determine that $\lim_{T \to 0}(T \frac{\partial S}{\partial T})_x = \lim_{T \to 0}(0 \times c_1) = 0$, where $\lim_{T \to 0}(\frac{\partial S}{\partial T})_x = c_1$, which is required to be a finite value to ensure that Eq. (31) holds. Moreover, one can derive $\lim_{T \to 0} \Delta S = \lim_{T \to 0} c_1 \Delta T = 0$, resulting in $\lim_{T \to 0} \Delta S = \lim_{T \to 0} (\Delta S)_T = 0$ because the thermodynamic process of a system at $T \to 0$ is isothermal. This shows that Eq. (12) can be directly derive from Eq. (16).

So far it has been proved through three different methods that Eqs. (12) and (16) may be mutually deducible and are equivalent. It is important that in the proof processes, the unattainability statement of absolute zero temperature is not involved.

## 9. The unattainability statement of absolute zero temperature

The unattainability statement of absolute zero temperature has appeared in textbooks for more than one century and may be stated as follows [3]: It is impossible by means of any procedure, no matter how idealized, to reduce the temperature of a system to the absolute zero in a finite number of steps. It can be mathematically expressed as [52]

$$-(\Delta T)/T < 1. \tag{32}$$

Because the inherent properties of the absolute temperature [2, 3, 5, 73], $-(\Delta T)$ is impossibly larger than $T$ for any adiabatic cooling process starting from an arbitrary low temperature. The lower the temperature in any cooling process is, the more difficult to cool it down. At present, spin gradient cooling [74] is permitted to reach the internal kinetic energy below 350 pK. By employing matter-wave lenses based on magnetic [75], electrostatic [76] forces make it possible to reduce the internal kinetic energy of a Bose-Einstein condensation system to about 50 pK [77]. Although the absolute zero temperature cannot be attained, it does not negate that the absolute zero temperature can be infinitely approached [42]. The above analyses show that the unattainability statement of absolute zero temperature can be viewed as a direct conclusion of temperature measurement.

Below, it will be discussed that the relation between the unattainability statement of absolute zero temperature and the Nernst statement or the heat capacity statement.

When a thermodynamic system evolves along a quasistatic reversible adiabatic process, the entropy $S$ remains unchanged. According to the Nernst equation, Eq. (29) can be simplified as [52, 78]

$$F(x',T_1) = F(x'',T_2). \tag{33}$$

It is seen from Eq. (33) that when $T_1 > 0$, $F(x',T_1) > 0$ and $T_2$ cannot attain zero, i.e., the unattainability of absolute zero temperature. It should be pointed out that in the whole derivative process, Eqs. (12) and (16) have been applied. This shows clearly that the unattainability statement of absolute zero temperature can be directly derived from Eq. (12) or (16), because both Eq. (12) and Eq. (16) are mutually deducible.

The second method is directly to generate the temperature-entropy diagram [3, 6, 7] of a thermodynamic system by using Eqs. (12) and (27), as shown in Fig. 10. It is directly seen from Fig.10 that the unattainability of absolute zero temperature in a finite number of steps can be derived from Eq. (12) or Eq. (16), because Eq. (27) includes Eq. (16).

The third method commonly used in thermodynamics [3], i.e., a proof by contradiction, is given as follows. If the Nernst statement is assumed not to be true, the first case is to assume $S(0,x'') > S(0,x')$ in Eq. (29), so that one can find a positive $T_1$ to ensure that $F(x',T_1)$ is small enough, $S(0,x'') - S(0,x') = F(x',T_1)$, and $T_2 = 0$; the second case is to assume $S(0,x'') < S(0,x')$ in Eq. (29), the reversible adiabatic process goes in reverse, a positive $T_2$ can be found to ensure that $F(x',T_2)$ is small enough, $S(0,x') - S(0,x'') = F(x'',T_2)$, and $T_1 = 0$. The above proofs show that only if the Nernst statement is not true, can the absolute zero temperature attain. It is proved once again that Eq. (12) or (16) can be used to derive the unattainability statement of absolute zero temperature, while no any additional assumptions are required.

Although that the unattainability statement of absolute zero temperature does not imply the Nernst equation has long been pointed out [22], there has been no consensus on this view. For example, in some textbooks [3, 46] and articles [79], Eq. (29) is directly used to prove that the Nernst theorem (i.e., the Nernst equation) can be derived from the unattainability statement of absolute zero temperature. It is natural because Eq. (16) has been applied to Eq. (29) and the Nernst equation can be directly derived from Eq. (16). It has been clearly pointed out in textbook [42] that in the process of the Nernst equation derived from the unattainability statement of absolute zero temperature, the vanishing heat capacity at absolute zero temperature must be used, which is a key point. However, this issue has not been noted in most of textbooks. Without the aid of Eq. (16), no mathematical connection can be established between the Nernst equation and the unattainability statement of absolute zero temperature, let alone that the former can be derived from the latter.

It should be pointed out that Fig.10 includes the contents of Eqs. (12) and (16). If Fig.10 is used to prove that the Nernst equation can be directly derived from the unattainability statement of absolute zero temperature, it will be a logical error similar to the description above.

The above analyses show that only when the substantive role of the heat capacity statement is clearly expounded, can the issue of whether both the Nernst statement and the unattainability statement of absolute zero temperature are equivalent or not be deeply recognized.

In fact, it is intuitively seen from the mathematical structures of Eqs. (12) and (32) that both the Nernst statement and the unattainability statement of absolute zero temperature cannot be equivalent. The former is an equation, and the latter is an inequality. It is also seen from the physical contents of Eqs. (12) and (32) that two statements mentioned above cannot be equivalent, because the Nernst equation may be used to discuss the performance of thermodynamic systems at $T = 0$, while the unattainability statement of absolute zero temperature has never been involved in the performance of thermodynamic systems at $T = 0$.

So far it has been expounded that the unattainability statement of absolute zero temperature is only a corollary of the Nernst statement or the heat capacity statement. Both the unattainability statement of absolute zero temperature and the Nernst statement are not equivalent. Of course, both the unattainability statement of absolute zero temperature and the heat capacity statement are also not equivalent.

**10. Two equivalent statements of the third law of thermodynamics**

It is clearly seen from the above analyses that the Nernst theorem and the unattainability statement of absolute zero temperature, which have appeared in textbooks [2, 3, 5-7] for more than one hundred years, are unsuitably taken as the statements of the third law of thermodynamics. The Nernst statement and the heat capacity statement can be used as two statements of the third law of thermodynamics, so that the core content of the third law of thermodynamics is a true reflection of the objective world, without any artificial additional assumptions. These two innovative equivalent statements will play an important role in improving the theoretical framework of classical thermodynamics. This discovery can effectively prevent some artificial assumptions into the third law of thermodynamics, and end the century-old debate about the statements of the third law of thermodynamics.

To date, only two equivalent statements of the third law of thermodynamics have been found, unlike the second law of thermodynamics, which can have many equivalent statements. Are there other equivalent statements for the third law of thermodynamics? This will be a new open question.

## 11. The applications of the third law of thermodynamics

Besides the unattainability statement of absolute zero temperature can be derived from the third law of thermodynamics, many important properties of thermodynamic systems at $T=0$ can be also derived directly from the third law of thermodynamics. For example, it can be directly deduced from Eq. (12) that the expansion coefficient, pressure coefficient, and zero-point heat of thermodynamic systems at $T=0$ are equal to zero.

The third law of thermodynamics can be directly used to discuss the fluctuations of thermodynamic systems at $T=0$. According to the thermodynamic method of fluctuations [80], the mean square fluctuations of the entropy and temperature for a thermodynamic system described by Eq. (1) are, respectively, given by [5, 81]

$$\overline{(\Delta S)^2} = kT(\partial S/\partial T)_Y = kC_Y \tag{34}$$

and

$$\overline{(\Delta T)^2} = kT(\partial T/\partial S)_y = kT^2/C_y, \tag{35}$$

where $k$ is the Boltzmann constant. It is seen from Eqs. (34) and (35) that the mean square fluctuations of the entropy and temperature of a thermodynamic system are, respectively, dependent on the heat capacities $C_Y$ and $C_y$ of the system. Because $\lim_{T \to 0} C_Y = 0$, Eq. (34) at $T \to 0$ is simplified as

$$\lim_{T \to 0} \overline{(\Delta S)^2} = 0. \tag{36}$$

It indicates clearly that the mean square fluctuation of the entropy of thermodynamic systems at $T \to 0$ is equal to zero. When $T \to 0$, the mean square fluctuation of the temperature of a thermodynamic system is different from that of the entropy and cannot be reduced to one case. It is directly dependent on the concrete circumstances that the heat capacity $C_y$ of the system tends to zero as the temperature approaches zero. For example, for the Fermi gas, $C_v \propto T$ [82, 83] and Eq. (35) at $T \to 0$ is simplified as

$$\lim_{T \to 0} \overline{(\Delta T)^2} = 0; \tag{37}$$

while for the photon gas, $C_v \propto T^3$, $U \propto T^4$ [82, 83], and Eq. (35) can be simplified as

$$\overline{(\Delta T)^2} \propto 1/T. \tag{38}$$

Eq. (38) indicates that when temperature is low enough, the fluctuation of the temperature of the system is not a small amount and the concept of thermal equilibrium is no longer applicable to such a system. When $T \to 0$, $U \to 0$, $\overline{(\Delta T)^2}$ tends to infinite, and the photo gas system goes to extinction. In a similar way, the fluctuations of the other parameters of thermodynamic systems at $T = 0$ can be also discussed.

The third law of thermodynamics can be also used to discuss the properties of thermodynamic systems at very low temperatures. For example, the third law of thermodynamics can determine which expressions of heat capacity may appear in ultra-low temperature thermodynamic systems and which expressions cannot appear in ultra-low temperature thermodynamic systems and are fictitious [54]. According to the third law of thermodynamics, we can determine which equations of state of thermodynamic systems are not suitable for ultra-low temperatures. For example, for dielectric materials satisfying the Curie law and the Curie-Weiss law [84-87], the heat capacities may not satisfy Eq. (16), and consequently, the third law of thermodynamics requires the Curie law and the Curie-Weiss law of dielectric materials to fail at ultra-low temperatures. All the heat capacities of classical gases do not satisfy Eq. (16). Thus, the equations of state of classical gases are incompatible with the third law of thermodynamics and cannot be used to discuss the properties of gases at ultra-low temperatures. In practice, for various systems whose heat capacity may be expressed as

$$C_x = aT^\lambda, \tag{39}$$

where $\lambda \leq 0$, the equations of state of these systems are not compatible with the third law of thermodynamics because Eq. (39) does not satisfy Eq. (16). This shows that the heat capacity of any thermodynamic system at ultra-low temperatures does not appear in the form of Eq. (39). When the heat capacity does not satisfy Eq. (16), it is possibly taken as the expression of a thermodynamic system at high or low temperatures but is impossibly the expression of a thermodynamic system at ultra-low temperatures. It is very interesting to note that one can mathematically devise a class of the expressions of the heat capacity satisfying Eq. (16), but they make the integral $\int_0^T C_x / T \, dT$ in Eq. (27) be divergent. Such heat capacities [53] are fictional and impossibly appear in practical ultra-low temperature systems [54]. Thus, it is unmeaningful to use a fictional heat capacity to discuss the performance of thermodynamic systems.

When the heat capacity of a thermodynamic system is given by

$$C_x = A_1 T^\chi + A_2 T^\delta + A_3 T^\gamma + \ldots, \tag{40}$$

where $\chi > 0$, $\chi < \delta < \gamma$, $A_i \geq 0$ (i=1,2,3,...), but $A_i (i=1,2,3,...)$ cannot equal zero at the same time, the equation of state of the system is compatible with the third law of thermodynamics and can be used to discuss the properties of the system at ultra-low temperatures because Eq. (40) satisfies Eq. (16). It can be seen from Eq. (40) that the simplest form satisfying Eq. (16) is given by

$$C_x = A T^\chi, \tag{41}$$

where $A$ is a finite value larger than zero, which depends on the concrete properties of a thermodynamic system but is independent of temperature. For example, quantum Bose gas, $C_v \propto T^{3/2}$; quantum Fermi gas, $C_v \propto T$; photon gas, $C_v \propto T^3$; electron gas in metal, $C_v \propto T$; and superflow $^4He$, $C_v \propto T^3$ [1, 42, 83, 88]. These heat capacities satisfy Eq. (41), and consequently, the corresponding equations of state are compatible with the third law of thermodynamics and can be used to discuss the properties of the systems described above at ultra-low temperatures.

The third law of thermodynamics is also widely applied to other disciplines. For example, it can be used to check whether a newly proposed statistical physical model is suitable for ultra-low temperatures, etc.

## 12. Conclusions

The Nernst equation can be obtained directly from the experimental data of chemical reactions at low temperatures, without the need of artificial additional assumptions, which should not be called the Nernst theorem, but rather the Nernst statement. The Nernst theorem should be withdrawn from thermodynamic theory.

The heat capacity statement proposed recently can be obtained directly from experimental data of heat capacity at low temperatures. It can be proved in theory that it is mutually derivable with the Nernst statement, and the two are equivalent.

It can be proved through several different ways that the unattainability statement of absolute zero temperature is only a corollary of the Nernst statement or the heat capacity statement, and is not equivalent to the Nernst statement or the heat capacity statement.

The Nernst theorem and the unattainability statement of absolute zero temperature, which have been used in textbooks for more than one century, are unsuitably taken as the statements of the third law of thermodynamics. The Nernst statement and the heat capacity statement are two equivalent statements of the third law of thermodynamics, which can solve the century-old controversy of the third law of thermodynamics and ensure the perfect building of classical thermodynamic theory. This is the breakthrough of a cross-century problematic issue in classical thermodynamic theory.

**Acknowledgements**

This work is supported by the National Natural Science Foundation (No. 12075197), People's Republic of China.

[1] H. B. Callen, Thermodynamics and an Introduction to Thermostatistics, 2$^{nd}$ ed., Wiley, New York, 1985.

[2] M. W. Zemansky and R. H. Dittman, Heat and Thermodynamics, 7$^{th}$ ed., McGraw-Hill, New York, 1977.

[3] J. S. Hsieh, Principles of Thermodynamics, McGraw-Hill, New York, 1975.

[4] J. Kestin, A Course in Thermodynamics, Vol 2, Hemisphere, New York, 1979.

[5] J. Chen, S. Su, and G. Su, Thermodynamics and Statistical Physics, 2$^{nd}$ ed., Science Press, Beijing, 2023.

[6] Z. C. Wang Thermodynamics and Statistical Physics, 5$^{rd}$ ed., Higher Education Publishing House, Beijing, 2013.

[7] Y. Xiong, Thermodynamics, 3th edn. People's Edu. Press, Peijing, 1979.

[8] E. A. Guggenheim, Thermodynamics, An Advanced Treatment for Chemists and Physicists, 5$^{th}$ ed., North-Holland, Amsterdam, 1967.

[9] E. G. Guggenheim, Thermodynamics, 4$^{th}$ ed., North-Holland, Amsterdam, 1959.

[10] M. W. Zemansky, Heat and Thermodynamics, 5$^{th}$ ed. McGraw-Hill, New York, 1968.

[11] J. Wilks, The Third Law of Thermodynamics, Oxford University Press, Oxford, 1963.

[12] A. Adamson, A Textbook of Physical Chemistry, Academic, New York, 1973.

[13] J. H. Noggle, Physical Chemistry, 2$^{nd}$ ed., Scott and Foresman Company Press, Glenview, 1989.

[14] A. Saggion, R. Faraldo, M. Pierno, Thermodynamics, Fundamental Principles and Applications, Springer, Cham, 2019.

[15] G. N. Hatsopoulos and J. H. Keenan, Principles of General Thermodynamics, John Wiley

& Sons, New York, 1965.

[16] R. Hasse, Physical Chemistry: An Advanced Treatise, Vol. 1/ Thermodynamics, ed. W. Jost, Academic, New York, 1971.

[17] I. N. Levine, Physical Chemistry, 2$^{nd}$ ed., McGraw-Hill, New York, 1983.

[18] P. T. Landsberg, Thermodynamics and Statistical Mechanics, Oxford, Oxford, 1978.

[19] J. C. Wheeler, Phys. Rev. A 43, 5289, 1991.

[20] J. C. Wheeler, Phys. Rev. A 45, 2637, 1992.

[21] G. N. Lewis and G. E. Gibson, J. Am. Chem. Soc. 42, 1529, 1920.

[22] P. T. Landsberg, Rev. Mod. Phys. 28, 363, 1957.

[23] P. T. Landsberg, Am. J. Phys. 65, 269, 1997.

[24] H. Reiss, Methods of Thermodynamics, Dover, New York, 1965.

[25] J. A. Beattie and I. Oppenheim, Principles of Thermodynamics, Elsevier, Amsterdam, 1979.

[26] Z. Yan and J. Chen, J. Phys. A 21, L707, 1988.

[27] P. T. Landsberg, J. Phys. A 22, 139, 1989.

[28] I. Oppenheim, J. Phys. A 22, 143, 1989.

[29] P. D. Gujrati, Phys. Lett. A, 151, 375, 1990.

[30] Z. Yan, J. Chen, and B. Andresen, Europhys. Lett. 55, 623, 2001.

[31] S. Blau and B. Halfpap, Am. J. Phys. 64, 13, 1996.

[32] F. C. Andrews, Thermodynamics: Principles and Applications, Wiley-Interscience, New York, 1971.

[33] J. Kestin and J. R. Dorfman, A Course in Statistical Thermodynamics, Academic Press,

New York, 1971.

[34] R. Kubo, Thermodynamics: An Advanced Course with Problems and Solutions, North-Holland, Amsterdam, 1968.

[35] S. Su, Y. Zhou, G. Su, and J. Chen, Mod. Phys. Lett. B 37, 2250214, (2023).

[36] H. Callen, Thermodynamics: An Introduction to the Physical Theories of Equilibrium Thermostatics and Irreversible Thermodynamics, John Wiley and Sons, New York, 1960.

[37] H. Struchtrup, Phys. Rev. Lett. 120, 250602, 2018.

[38] R. B. Griffiths, J. Math. Phys. 5, 1215, 1964.

[39] L. Galgani and A. Scotti, Physica 40, 150, 1968.

[40] L. Galgani and A. Scotti, Physica, 42, 242, 1969.

[41] L. Galgani and A. Scotti, Pure Appl. Chem. 22, 229-236, 1970.

[42] Z. Lin, Thermodynamics and Statistical Physics, Peking University Press, Beijing, 2007.

[43] A. Rex and C. B. P. Finn, Finn's Thermal Physics, 3rd ed., CRC Press, Milton, 2017.

[44] P. Coffey, Hist. Stud. Phys. Biol. Sci. 36, 365, 2006.

[45] J. Ganguly, Thermodynamics in Earth and Planetary Sciences, 2nd ed., Springer Nature: Switzerland AG, 2020.

[46] Z. Wang, Thermodynamics, Higher Education Publishing House, Beijing, 1955.

[47] W. Nernst, Ber. Kgl. Pr. Akad. Wiss. 52, 933, 1906.

[48] J. Wilks, The Third Law of Thermodynamics, Oxford University Press, Oxford, 1961.

[49] S. Su, J. Y. Chen, and J. Chen, College Phys. 23, in press, 2024.

[50] S. Su, S. Xia, T. Liang, and J. Chen, Mod. Phys. Lett. B 38, 2450115, 2024.

[51] J. Bao, Brief Course on Thermodynamics and Statistical Physics, Higher Education Press,

Beijing, 2011.

[52] X. Chen, Y. Zhou, and J. Chen, Chin. Phys. B. 2004, doi.org/10.1088/ 1674-1056/ad39c8.

[53] D. C. Mattis, Statistical Mechanics Made Simple, World Scientific Publishing Co., Singapore, 2003.

[54] S. Su, J. Wang, X. Chen, and J. Chen, Mod. Phys. Lett. B 37, 2350255, 2023.

[55] H. O'neal and N. E. Phillips, Phys. Rev. 137, A748, 1965.

[56] D. L. Martin, Phys. Rev. 170, 650, 1968.

[57] M. Barucci, S. Di Renzone, E. Olivieri, L. Risegari, and G. Ventura, Cryogenics 46, 767, 2006.

[58] O. V. Lounasmaa, Experimental Principles and Methods Below 1 K. Academic Press, London, 1974.

[59] K. M. Al-Shibani and O. A. Sacli, Phys. Stat. Sol. (b) 163, 99, 1991.

[60] M. Barucci, C. Ligi, L. Lolli, A. Marini, V. Martelli, L. Risegari, and G. Ventura, Physica B 405, 1452, 2010.

[61] H. R. O'Neal and N. E. Phillips, Phys. Rev. 137, A748, 1965.

[62] A. Nittke, M. Scherl, P. Esquinazi, W. Lorenz, J. Li, and F. Pobell, J. Low Temp. Phys. 98, 517, 1995.

[63] R. B. Stephens, Phys. Rev. B 8, 2896, 1973.

[64] M. Barucci, E. Gottardi, E. Olivieri, E. Pasca, L. Risegari, and G. Ventura, Cryogenics 42, 551, 2002.

[65] D. S. Greywall, Phys Rev B, 27 2747, 1983.

[66] W. F. Brinkman and S. Engelsberg, Phys Rev 169, 417, 1968.
39

[67] D. Szewczyk and M. A. Ramos, Crystals 12, 1774, 2022.

[68] R. C. Zeller and R. O. Pohl, Phys. Rev. B, 4, 2029, 1971.

[69] S. Su and J. Chen, Mod. Phys. Lett. A 37, 2250246, 2023.

[70] A. Munster, Statistical Thermodynamics, Springer, Berlin, 1974.

[71] N. F. Ramsey, Phys. Rev. 103, 20, 1956.

[72] S. Su and J. Chen, College Phys. 23, 2024, DOI: 10.16854/j.cnki.1000-0712.

[73] Z. Yan, J. Xiamen Univ. 21, 402 , 1982.

[74] P. Medley, D. M. Weld, H. Miyake, D. E. Pritchard, and W. Ketterle, Phys. Rev. Lett. 106 195301, 2011.

[75] H. Müntinga, H. Ahlers, M. Krutzik, A. Wenzlawski, S. Arnold, Becker D, et al. Phys. Rev. Lett. 110, 093602, 2013.

[76] J. G. Kalnins, J. M. Amini, and H. Gould, Phys. Rev. A 72, 043406, 2005.

[77] T. Kovachy, J. M. Hogan, A. Sugarbaker, S. M. Dickerson, C. A. Donnelly, C. Overstreet, and M. A. Kasevich, Phys. Rev. Lett. 114, 143004, 2015.

[78] F. W. Sears and G. L. Salinger, Thermodynamics, Kinetic Theory, and Statistical Thermodynamics, 3$^{rd}$ ed., Addison-Wesley, Bostom, 1977.

[79] T. D. Kieu, Phys. Lett. A 383, 125848, 2019.

[80] Z. Yan and J. Chen, J. Chem. Phys. 96, 3170, 1992.

[81] Z. Yan, College Phys. 6(9), 14, 1987.

[82] K. Huang, Statistical Physics, 2$^{th}$ ed., John Wiley $ Sons, NewYork, 1987.

[83] R. K. Pathria and P. D. Beale, Statistical Mechanics, 3$^{rd}$ ed., Elsevier, Amsterdam, 2011.

[84] J. C. Andreson, Dielectrics, Chapman and Hall, London, 1963.

[85] M. E. Lines and A. M. Glass, Principles and Applications of Ferroelectrics and Related Materials, Clarendon Press, Oxford, 1977.

[86] R. Coelhe, Physics of Dielectrics for the Engineer, Elsevier Scientific, Amsterdam, 1979.

[87] J. Wang and J. Chen, Appl. Energy, 72, 495, 2002.

[88] L. D. Landau, L. P. Pitaevskii, and E. M. Lifshitz, Statistical Physics, Pergamon Press, Oxford, 1980.